\documentclass[fleqn,10pt]{wlscirep}
\usepackage[utf8]{inputenc}
\usepackage[T1]{fontenc}
\newcommand{\tth}{$^{\text{th}}$ }
\title{A Machine-Learned Spin-Lattice Potential for Dynamic Simulations of Defective Magnetic Iron} 
\author[1,*]{Jacob B. J. Chapman}
\author[1,+]{Pui-Wai Ma}
\affil[1]{United Kingdom Atomic Energy Authority, Culham Science Centre, Abingdon, Oxfordshire, OX14 3DB, United Kingdom}

\affil[*]{Jacob.Chapman@ukaea.uk}
\affil[+]{Leo.Ma@ukaea.uk}

\begin{abstract}
A machine-learned spin-lattice interatomic potential (MSLP) for magnetic iron is developed and applied to mesoscopic scale defects. It is achieved by augmenting a spin-lattice Hamiltonian with a neural network term trained to descriptors representing a mix of local atomic configuration and magnetic environments. It reproduces the cohesive energy of BCC and FCC phases with various magnetic states. It predicts the formation energy and complex magnetic structure of point defects in quantitative agreement with density functional theory (DFT) including the reversal and quenching of magnetic moments near the core of defects. The Curie temperature is calculated through spin-lattice dynamics showing good computational stability at high temperature. The potential is applied to study magnetic fluctuations near sizable dislocation loops. The MSLP transcends current treatments using DFT and molecular dynamics, and surpasses other spin-lattice potentials that only treat near-perfect crystal cases.
\end{abstract}
\begin{document}

\flushbottom
\maketitle

\thispagestyle{empty}


\section*{Introduction}

The success of density functional theory (DFT) \cite{Hohenberg_prb_1964,Kohn_pra_1965} has drastically advanced the scientific and technological aspects of materials development due to its unprecedented predictive power at a modest computational cost. However, the order $O(n^3)$ scalability of DFT calculations, where $n$ is the number of electrons, has severely limited the simulation box size and time scale. Machine-learned potentials (MP) have demonstrated their ability to perform scalable atomic scale simulations with DFT accuracy using only a fraction of its computational requirements \cite{Goryaeva_cms_2019}. Since the seminal work of Behler and Parrinello \cite{Behler_jcp_2011}, who introduced the concept of invariant descriptors to represent local chemical environment, a range of MP based on kernel methods \cite{Bartok_prl_2010,Bartok_prb_2013} and network networks \cite{behler_jcp_2016,Cubuk_prl_2015,Cooper_npj_2020} have been developed and applied to investigate real physical problems. 

Spin-polarized and non-collinear magnetism are well established extensions of DFT for magnetic materials but their results are valid only for the electronic ground state. Attempts to mimic magnetic excitation by coupling spin dynamics to constrained non-collinear calculations have been made \cite{antropov_prb_1996,Hellsvik_PRL_2019}. However, the limitations of the DFT method on the simulation box size has yet to be overcome. In addition the effects of magnetic excitation and their interaction with atomic trajectories are irreconcilable within the framework of classical molecular dynamics (MD) \cite{Ma_Handbook_2020}.

Nevertheless, magnetic effects cannot be ignored in many situations. In magnetic iron, the bcc-fcc and fcc-bcc phase transitions at 1185K and 1667K, respectively, are due to the competing phonon and magnon free energies \cite{ma_prb_2017,Kormann_prb_2012,Kormann_prl_2014,Hasegawa_PRL_1983,Lavrentiev_PRB_2010}. The softening of tetragonal shear modulus $C'$ near the Curie temperature $T_C$ \cite{Hasegawa_JPhysF_1985,Dever_JAP_1972} and stability of anomalous $\langle 110\rangle$ dumbbell self-interstitial atom (SIA) configurations \cite{nguyen-manh_prb_2006,derlet_prb_2007,chapman_prb_2020} are also believed to be magnetically driven. Itinerant ferromagnetism, in the form of increased magnitudes of magnetic moment (MM), have been linked to the stability of grain boundaries and intergranular cohesion \cite{ye_silleten_prl_1998}.

Spin-lattice dynamics (SLD) \cite{ma_prb_2008} was developed to treat both spin (magnetic) and lattice degrees of freedom within a unified framework. SLD is a general framework similar to MD and applicable to any arbitrary atomic scale Hamiltonian. The latest development on the Langevin spin equation of motion \cite{ma_prb_2012} allows simultaneous treatment of both the rotation and longitudinal fluctuation (magnitude) of MM. In most other studies the magnitude of MM are assumed to be fixed \cite{Tranchida_jcompphys_2018,Mudrick_2017,evans_jpcm_2014} or have been performed on a fixed lattice \cite{Chapman_prb_2019,malerba_nme_2021}. While SLD has been used to investigate a variety of microscopic dynamic effects in iron \cite{ma_prb_2008,ma_prb_2017,wen_jnucmater_2013,Mudrick_2017,evans_jpcm_2014,Nikolov_npj_2021,Tranchida_jcompphys_2018}, there is still not a spin-lattice potential (SLP) capable of simultaneously modelling mechanical deformations, magnetic fluctuations and defect properties \cite{Ma_Handbook_2020}. 

The difficulty of developing SLP is two-fold. First, a SLP has double the degrees of freedoms ($6N$) of a conventional MD potential ($3N$), where $N$ is the number of atoms. A substantial amount of extra data is required for potential fitting for each extra degree of freedom, drastically expending the representable phase space. Recent data-driven techniques can aid in parameter optimisation for such cases \cite{Nikolov_npj_2021}. Second, potentials that adopt the Heisenberg or Heisenberg-Landau functional form in various studies \cite{chapman_prb_2020} are shown to be too restrictive to near-perfect crystal cases. A good functional form that is applicable to both perfect and defective configurations is yet to be derived. 

MP for SLD that goes beyond the need of a well defined functional form could be a viable solution. While the number of MP for iron has rapidly increased over the past decade \cite{dragoni_prm_2018,Goryaeva_cms_2019,goryava_prm_2021,wang_cms_2022}, applications including explicit spin degrees of freedoms are very limited. Recently, Nikolov \textit{et al.} \cite{Nikolov_npj_2021} produced a machine-learned spectral neighbor analysis potential. Since they kept using the Heisenberg functional form, the potential does not consider the change of the magnitude of MM due to thermal excitation or the change of local atomic environment. Novikov \textit{et al.} \cite{Novikov_natcompmat_2022} developed a moment tensor SLP that includes longitudinal fluctuation, but they limited their approach to collinear configurations near perfect crystal structures. 

In this paper, we show that our newly developed machine-learned spin-lattice potential (MSLP), based on an alternative approach, is capable of describing the complex magnetic states at highly deformed as well as near-perfect configurations. Our MSLP for iron has good quantitative agreement with DFT data and good computation stability at high temperature simulations. The calculated $T_C$ is also in good agreement with experimental value. We also apply the MSLP to study the magnetic effect of mesoscopic scale dislocation loops in iron at finite temperature, which cannot be achieved using DFT or MD, or using other available MSLP.

\section*{Results}

\subsection*{Magnetic states in BCC and FCC structures}

\begin{table*}[]
\caption{\label{tab:cg} The equilibrium lattice constant $a_0$, the magnitude of spontaneous magnetic moment $|\mathbf{M}|$ and the relative energy difference with respect to the BCC ground state $\Delta E$ calculated using our machine-learned spin-lattice potential (MSLP) for iron at non-magnetic (NM), ferromagnetic (FM), single layer antiferromagnetic (SL-AFM), and double layer antiferromagnetic (DL-AFM) states in BCC and FCC structures. DFT calculations using VASP and OpenMX are shown for comparison. Details are in Supplementary Materials.}
\begin{tabular}{|l|l|ccc|ccc|ccc|}
\hline\hline 
                     \multicolumn{2}{|c|}{}       & \multicolumn{3}{c|}{MSLP}                      & \multicolumn{3}{c|}{DFT (VASP)}      & \multicolumn{3}{c|}{DFT (OpenMX)}  \\
                      \multicolumn{2}{|c|}{}        & $a_0$    & $|\mathbf{M}|$   & $\Delta E$ & $a_0$  & $|\mathbf{M}|$ & $\Delta E$   & $a_0$ & $|\mathbf{M}|$ & $\Delta E$ \\ 
                      \multicolumn{2}{|c|}{}        & ~~~~(\AA)~~~~~      & ($\mu_B$)      & (eV/atom)  & ~~~~(\AA)~~~~  & ($\mu_B$)      & (eV/atom)    & ~~~~(\AA)~~~~      & ($\mu_B$)      & (eV/atom)  \\ \hline
                     & FM     & 2.817    & 2.16             &            & 2.831  & 2.19           &              & 2.842 & 2.25 &  \\
BCC                  & SL-AFM & 2.824    & 1.54             & 0.36       & 2.800  & 1.34           &  0.46        & && \\
                     & NM     & 2.753    & 0.00             & 0.42       & 2.764  & 0.00           &  0.47        & 2.766 & 0.00 & 0.56 \\ \hline
                     & DL-AFM & 3.470    & 2.08             & 0.08       & 3.466  & 2.04           & 0.08         & 3.476 & 2.38 & 0.10 \\
FCC                  & SL-AFM & 3.494    & 0.96             & 0.16       & 3.494  & 1.30           & 0.12         & 3.435 & 2.00 & 0.13\\
                     & FM     & 3.47     & 1.03             & 0.15       & 3.50   & 1.00           & 0.16         & 3.648 & 2.63 & 0.12 \\
                     & NM     & 3.428    & 0.00             & 0.18       & 3.456  & 0.00           & 0.16         & 3.462 & 0.00 & 0.25  \\
\hline \hline
\end{tabular}
\end{table*}

We investigated an essential feature being necessary for a MSLP for iron, which is the relative stability of various magnetic states in BCC and FCC structures. We initialized the ferromagnetic (FM), single-layer antiferromagnetic (SL-AFM), and non-magnetic (NM) states in both BCC and FCC structures, and additionally the double-layer antiferromagnetic (DL-AFM) state in FCC. We relaxed the simulation box and MM using conjugate gradient method, but with a small mixing step, to ensure the relaxation would stop at local minimum. Table \ref{tab:cg} summarizes our results. It shows the equilibrium lattice constant $a_0$, the magnitude of spontaneous MM $|\mathbf{M}|$ and the relative energy difference with respect to the BCC ground state $\Delta E$. DFT data calculated using both VASP \cite{KressePRB1993,KressePRB1994,KressePRB1996,KresseCMS1996} and OpenMX \cite{openmx} packages are shown for comparison.

FM BCC is the most stable state. There is small underestimation of the $a_0$ (-0.5\%) and $|\mathbf{M}|$ (-1.4\%) compared to VASP data. The DL-AFM is the lowest energy collinear state in FCC, which is 80 meV/atom higher than the FM BCC phase. The energy of other magnetic states are also in quantitative agreement with DFT data. The NM FCC was shown to have free energy lower than NM BCC at all temperatures \cite{ma_prb_2017}. It is magnetism that stabilizes the BCC structure \cite{ma_prb_2017,Kormann_prb_2012,Kormann_prl_2014,Hasegawa_PRL_1983,Lavrentiev_PRB_2010}. Our MSLP reproduces this phenomenon.

In BCC iron, the formation of spontaneous MM reduces the energy by 0.42 eV/atom. By varying the magnitude of MM we can plot the Landau-functional-like energy well (Supplementary Materials). The position and depth of the minimum for FM state is well reproduced resulting in accurate properties of the FM BCC phase. However, a small discrepancy on the profile of the curve compared to DFT data can be observed for small MM. We note our MSLP predicts a different order of stability of the magnetic states in FCC relative to VASP data, where a low spin FM state has slightly lower energy than the SL-AFM. On the other hand, DFT data from OpenMX predicts the same order of stability as our potential. This highlights the complexity of the potential energy surface of iron where the relative stability of magnetic states is in the order of 0.01 eV.

Our MSLP produced various magnetic states quantitatively as good as the moment tensor SLP developed recently by Novikov \textit{et al.} \cite{Novikov_natcompmat_2022}, which is valid only near-perfect crystal collinear regime. More details on the comparison of energies, forces, stresses and effective magnetic fields with respect to DFT data are in Supplementary Materials.  

\subsection*{Finite temperature properties: lattice constant and Curie temperature}

\begin{figure}
\includegraphics[width=16.cm]{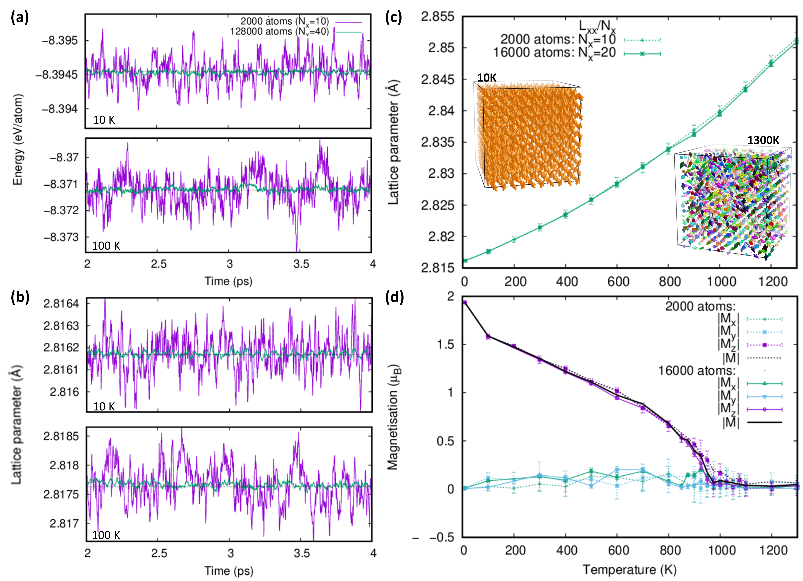}
\caption{\label{fig:lvt} Dynamic stability and properties of the machine-learned spin-lattice dynamics potential (MSLP) for iron. Comparison of instantaneous (a) total energies per atom and (b) lattice parameter per unit cell between 2,000 atom ($10\times 10\times 10$) and 128,000 atom ($40\times 40\times 40$) simulation cells at 10 K and 100 K. (c) Lattice constants and (d) magnetization ($|\mathbf{M}|=|\sum_i \mathbf{M}_i |/N$) calculated using our MSLP for iron using a simulation boxes containing 2,000 and 16,000 atoms. Standard deviation of data are shown as error bars. Details are in Supplementary materials. Subfigures in (c) show snapshots of the ferromagnetic arrangement of magnetic moments at 10 K and the paramagnetic state at 1300 K.
}
\end{figure}

The main purpose of developing a MSLP is to perform dynamic simulations at finite temperature and observe the time evolution of a system. We implemented our MSLP into SPILADY \cite{spilady}. It allows SLD to be performed with longitudinal fluctuations of MM \cite{ma_prb_2008,ma_prb_2011,ma_prb_2012} which is a unique feature of the code and a fundamental concept built into the MSLP.  

The initial calculations prove dynamic stability. In SLD it is important that the potential energy surface is smooth and continuous because both atomic forces and effective magnetic fields are derivatives of the Hamiltonian. A small abnormality may generate unexpected artefacts such as large forces or magnetic fields that destroy the system. Figure \ref{fig:lvt}a shows the total energies of 2,000 and 128,000 atom FM BCC Fe SLD simulations in NPT ensembles. The magnitude of energy fluctuation is inversely proportional to the number of particles. The average energy of both size runs are equivalent with no evidence of drift. Figure \ref{fig:lvt}b shows the lattice parameters of the same calculations confirming the consistency of the potential with simulation size. Scalability is important since simulations of the order 10$^5$ and larger are beyond the current capability of DFT studies of metallic systems.

We examined the change of lattice constants and $T_C$ of our MSLP for BCC iron. We created cubic simulation boxes containing 2,000 and 16,000 atoms. Fig. \ref{fig:lvt}c shows the lattice constant, which is calculated from the time average of the linear dimension of a varying simulation box with pressure set to zero. The lattice constant monotonically increases with a smooth slope as temperature increases. It is generally underestimated but comparable to other MD potentials\cite{proville_natmat_2012}. The standard deviation, which is shown as the error bar, remains small even at high temperature showing good stability of our potential.

$T_C$ is an unique indicator of a SLP. BCC iron undergoes ferromagnetic to paramagnetic phase transition at 1043K \cite{Lavrentiev_JPhysCM_2012}. Fig. \ref{fig:lvt}d shows the calculated magnetization $|\mathbf{M}|=|\sum_i \mathbf{M}_i|/N$. Calculations were performed in a smaller step of 25K near the $T_C$. The calculated $T_C$ is around 900K, which is in reasonable agreement with experiment.

\subsection*{Point Defects: Self-interstitial Atom and Vacancy}

\begin{table}[]
\caption{\label{tab:sia} Magnetic moments in the vicinity of a $\langle 110 \rangle$ dumbbell self-interstitial atom and vacancy configurations. The core, compressive and tensile sites refer to the positions defined in the inset of Fig. \ref{fig:sia}. They are all in unit of Bohr magnetons ($\mu_B$). Bulk-like refers to atoms far from the defect core.}
\begin{tabular}{|l|l|cccc|}
\hline \hline 
 Defect                             & Site        & MSLP & VASP-PAW     & OpenMX \cite{chapman_prb_2020}   & VASP-USSP\cite{olsson_prb_2007}               \\ \hline\hline
                                    & Core        & -0.28 & -0.21        & -0.30         & -0.18                   \\
 $\langle 110 \rangle_{\text{DB}}$  & Compressive & 1.70  & 1.66         & 1.87          & 1.52                    \\
                                    & Tensile     & 2.31  & 2.37         & 2.45          & 2.30                    \\ \hline
                                    & 1NN         & 2.23  & 2.43         & 2.53          & 2.70                    \\
Vacancy                     & 2NN         & 2.08  & 2.08         & 2.13          & 2.41                    \\
                                    & 3NN         & 2.10  & 2.21         & 2.24          &                         \\ \hline 
                                    & Bulk-like   & 2.11  & 2.19         & 2.22          & 2.52                    \\ \hline\hline
\end{tabular}
\end{table}

\begin{figure*}[]
\includegraphics[width=16cm]{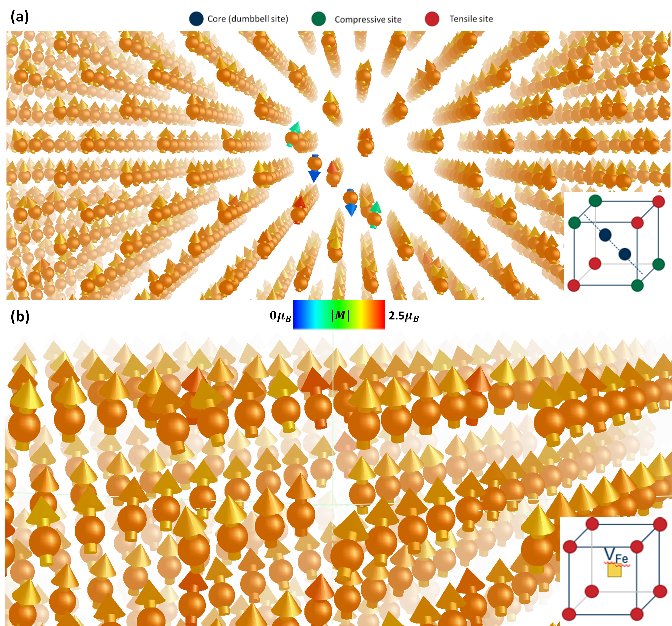}
\caption{\label{fig:sia} A snapshot of (a) a $\langle 110 \rangle$ dumbbell self-interstitial atom configuration and (b) a mono-vacancy in iron at 10K. Magnitude of magnetic moments are indicated according to the colour bar. The insets shows the schematic configurations. For the SIA, the schematic indicates the core (blue), compressive (green) and tensile (red) sites.}
\end{figure*}

DFT calculations show in highly distorted lattice structures, complex magnetic configurations can be observed. Existing SLP \cite{Nikolov_npj_2021,Novikov_natcompmat_2022,ma_prb_2017,ma_prb_2008} remain incapable of capturing such phenomena even for point defects. The magnitude of MM near the defect core can be suppressed or even in reverse alignment with respect to the bulk \cite{olsson_prb_2007}. Models that only adopt the Heisenberg Hamiltonian cannot produce physically correct point defect migration as the model does not allow change of magnitude of MM according to the change of local electronic structure \cite{chapman_prb_2020}. Additional Landau terms in the Hamiltonian which are functions of local environment may be a solution but correct treatment of itinerant properties remain unresolved \cite{chapman_prb_2020}.

In BCC iron, the most stable SIA configuration is a $\langle 110\rangle$ dumbbell configuration \cite{nguyen-manh_prb_2006,derlet_prb_2007}. Using our MSLP, we performed annealing simulations with initial configurations including either a $\langle 110 \rangle$ or $\langle 111 \rangle$ dumbbell, in a cell with 2001 atoms. The cells were initially thermalised to 10K and gradually decreased to 0K for 5ps. Both SIA configurations relaxed to maintain/form a $\langle 110 \rangle$ dumbbell. We can understand this through nudged elastic band DFT calculations which show no intermediary energy barriers across the migration pathway between the $\langle 111 \rangle$ and $\langle 110 \rangle$ SIA configurations (see Supplementary Materials). A $\langle 111 \rangle$ SIA configuration will inevitably relax to a $\langle 110 \rangle$ dumbbell when small perturbations exists. A snapshot of SLD simulation of the $\langle 110 \rangle$ configuration at 10K is shown in Fig. \ref{fig:sia}a. The MMs were plotted with unit magnitude for ease of viewing. Their magnitudes are represented by colour. 

We examined the magnetic configuration in the core of a $\langle 110 \rangle$ dumbbell. The MM in and surrounding the core are listed in Table \ref{tab:sia}. It shows very good agreement in comparison to DFT calculations: VASP with PAW pseudopotentials (current work), VASP with ultrasoft pseudopotentials (USPP) \cite{olsson_prb_2007}, and OpenMX \cite{chapman_prb_2020}. The magnitude of the MM within the defect core are larger than the VASP-PAW data that the potential was trained to, but is similar to those produced by OpenMX. Likewise, for the tensile site the MMs predicted by the MSLP are smaller than VASP-PAW data but are comparable to VASP-USSP data. Generally, the MMs are reproduced in quantitative agreement with DFT calculations. In the core of the interstitial defect, magnitudes of MMs are approximately 1/10\tth of bulk and in anti-alignment to the bulk. Enhanced magnitudes can be observed on the tensile sites and slightly reduced magnitudes on the compressed sites. In additional to the most stable configuration, our MSLP reproduced the correct order of stability of SIA, i.e. the formation energy of $\langle 110 \rangle$ < tetrahedral < $\langle 111 \rangle$ < $\langle 100 \rangle$ < octhahedral (see Supplementary Materials).

Another point defect that we explored is the mono-vacancy ($V_{\text{Fe}}$). Annealing simulations were performed using a 1999 atoms cell containing a single vacancy. Table \ref{tab:sia} shows the calculated values of MMs in the vicinity of the defect site. Fig. \ref{fig:sia}b shows a snapshot of the system near the vacancy during dynamics. DFT calculations indicate that the MM directly adjacent to a vacancy are larger. This occurs due to the increased volume to which their moments can relax.  The magnitude of the MMs in the 1$^\text{st}$ nearest neighbour (NN) sites are approximately 11\%, 14\% and 6.7\% larger for the VASP-PAW, OpenMX and VASP-USSP calculations. The increase is only 5.6\% using the MSLP. Conversely, the MMs of the 2$^\text{nd}$ NN to the vacancy have reduced magnitudes. Our potential predicts a reduced MM relative to bulk in line with DFT calculations, but the proportion is diminished. By the 3$^\text{rd}$ NN sites, the MMs are \textit{bulk-like} in all cases. Our MSLP predicted the correct trend of the changes, but generally gives a smaller value.

\subsection*{Extended defects: prismatic dislocation loops}

\begin{figure*}
\includegraphics[width=15cm]{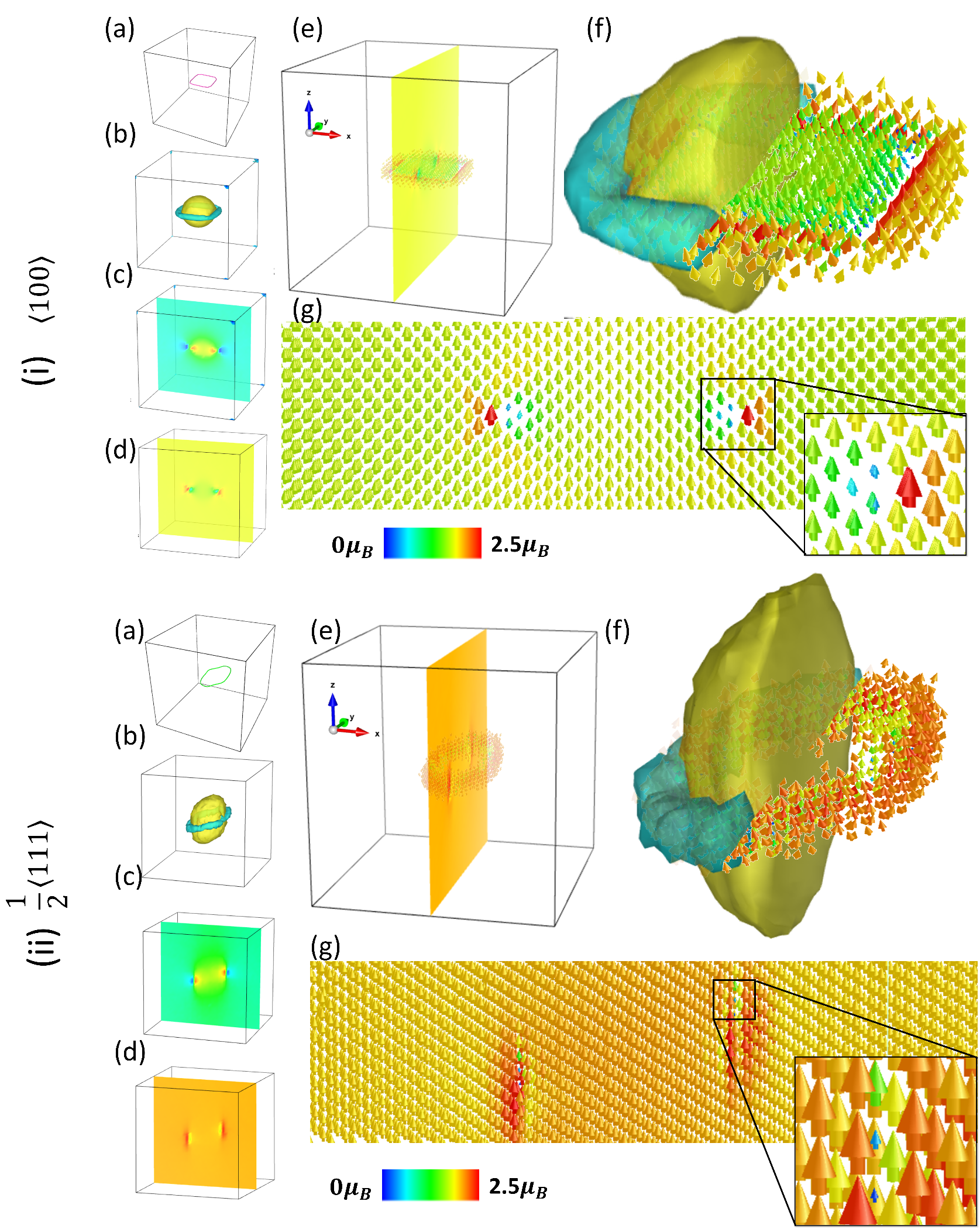}
\caption{\label{fig:sia10K} 
Spin-lattice dynamics simulations performed at 10~K for the (i) $\langle 100 \rangle$ and (ii) $\frac{1}{2}\langle 111 \rangle$ interstitial dislocation loops. (a) Dislocation identified using the dislocation extraction algorithm (DXA). (b) The compressive and tensile stresses caused by the dislocations are shown via the isosurface of $\text{Tr}(\sigma_{ij}^k)$, where $\sigma_{ij}^k$ is the Virial stress tensor of atom $k$. (c) 2D contour plot of $\text{Tr}(\sigma_{ij}^k)$ through the (010) or $(\bar{1}2\bar{1})$ plane bisecting the centre of loop. (d) 2D contour plot of the MM magnitudes on the same plane as (d). (e) Contour plot of MM magnitudes overlayed with a sample of the instantaneous MM vectors for atoms near the loop. (f) Overlay of magnetic moments near the loop with the stress tensor isosurface showing relation between compressive (blue) and tensile (yellow) regions with large (red/orange) and small (blue/green) magnetic moments (see colourbar). (g) Vector field of the MMs near the dislocation loops highlighting the core of the dislocations with suppressed magnetic moments and the enhanced moments directly adjacent. 
}
\end{figure*}

\begin{figure*}
\includegraphics[width=15cm]{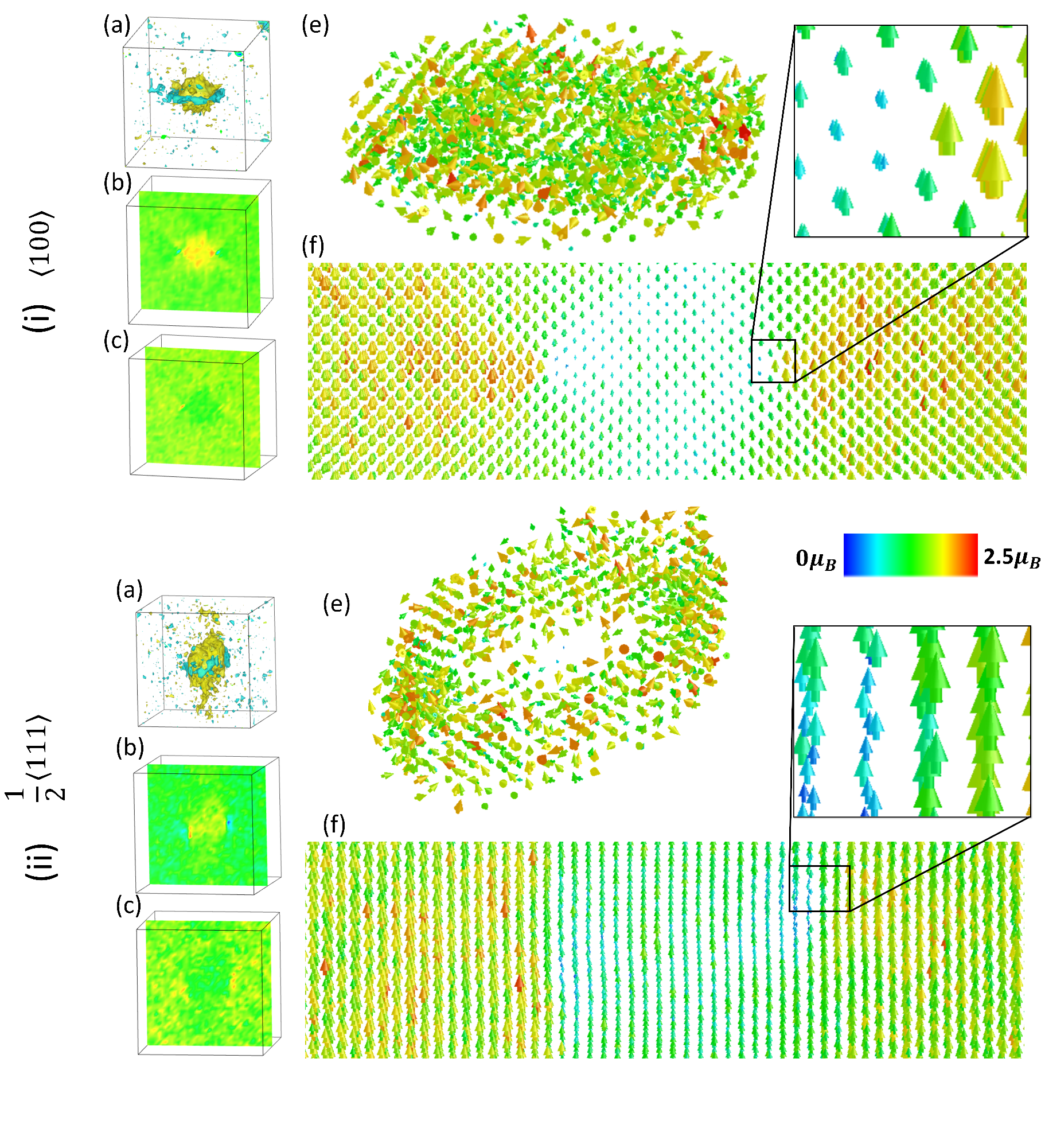}
\caption{\label{fig:sia800K} 
Spin-lattice dynamics simulations performed at 800~K for the (i) $\langle 100 \rangle$ and (ii) $\frac{1}{2}\langle 111 \rangle$ interstitial dislocation loops.  (a) The compressive and tensile stresses caused by the dislocations are shown via the isosurface of $\text{Tr}(\langle \sigma^k_{ij}(t)\rangle)$, where $\langle \sigma^k_{ij}(t)\rangle$ is the time averaged Virial stress tensor of atom $k$. (b) 2D contour plot of $\text{Tr}(\langle \sigma^k_{ij}(t)\rangle)$ through the (010) or $(\bar{1}2\bar{1})$ plane bisecting the centre of loop. (c) 2D contour plot of the MM magnitudes on the same plane as (b). (d) Sample of the instantaneous magnetic moments during dynamics of atoms in and near the dislocation loop. (e) Vector field representing the averaged magnetic moments near the loop.
}
\end{figure*}

We applied our MSLP to sizable systems that cannot be addressed by DFT. We constructed two simulation cells which are pre-relaxed using the Malerba 2010 Fe potential\cite{malerba_jnm_2010} through the conjugate gradient implementation in LAMMPS. 
In the first cell we created a square SIA loop with $\mathbf{b}=a_0[001]$ consisting of 265 atoms in a box containing 128,265 atoms. In the second, a circular SIA loop with $\mathbf{b}=\frac{a_0}{2}[111]$ was constructed with 261 atoms in a box containing 139,287 atoms. The relaxed prismatic dislocation loops identified using the dislocation extraction algorithm (DXA) are shown in Fig. \ref{fig:sia10K}(i)a for the $\langle 100 \rangle$ loop and \ref{fig:sia10K}(ii)a  for the $\frac{1}{2}\langle 111 \rangle $ loop.

We chose these dislocation loops as representative examples because both kinds of loop can be experimentally observed in $\alpha$-iron. Iron is known to be anomalous, forming $\langle 100 \rangle$-type prismatic edge dislocations at temperatures above 550$^{\circ}$C\cite{masters_nature_2963,little_jmicro_1973} despite the isotropic elasticity favoring dislocation loops with smaller Burgers vectors such as $\frac{1}{2}\langle 111\rangle$. Analytic linear elasticity solution suggests the softening of $C'$, which is a magnetic effect, accounts for the observation of square $\langle 100 \rangle$ loops at high temperature \cite{dudarev_prl_2008,dudarev_jnm_2009}. 

The MSLP offers analysis of magnetic excitation in the vicinity of these extended defects for the first time. As such, we performed SLD calculations in NPT ensembles at both 10K and 800K using the MSLP, evaluating the local stress and magnetic configurations of both loop types. Data for the $\langle 100 \rangle$ prismatic loop at 10~K are shown in subfigure~\ref{fig:sia10K}(i) whereas the $\frac{1}{2}\langle 111 \rangle $ loop results are in subfigure~\ref{fig:sia10K}(ii). For both loop types we present positive and negative isosurfaces of the stress field introduced by the defects. Specifically, we evaluated $\text{Tr}(\sigma^k_{ij})$, where $\sigma^k_{ij}$ is the Virial stress tensor of atom $k$ computed using our MSLP.

Yellow/blue isosurfaces show the compressed/tensile regions where atoms contribute $\text{Tr}(\sigma^k_{ij})$=$\pm$0.017 GPa to the stress for the $\langle 100\rangle$ loop and $\text{Tr}(\sigma^k_{ij})$=$\pm$0.024 GPa for the $\frac{1}{2}\langle 111\rangle$ loop. (c) and (d) show the contour maps of $\text{Tr}(\sigma^k_{ij})$ and MM magnitudes on a (100) or $(\bar{1}2\bar{1})$ plane intersecting the centre of the dislocation loops. (e) shows the MM vectors of atoms near the loop superimposed on the contour map. To visually compare the stress and magnetic configuration we present an overlay of the stress isosurface with a snapshot of the non-collinear MMs. The MMs are coloured according to their magnitude, with red/orange hues representing oversized moments, blue/dark-green hues for small and green moments for bulk-like magnitudes ($|\mathbf{M}|\approx 2.0\mu_B$). Vector fields of the time averaged moments with the same colour scheme are presented in (g) highlighting the MMs in the core region of the dislocation. A strong correlation is evident between the local stress and the MMs. Regions under tension, which have a comparatively larger volumes per atom, cause MMs to increase relative to their bulk value. On the other hand, regions which are compressed result in reduced magnitudes of the MMs. This is most evident in the core of the dislocation line. 

Data pertaining to the $\langle 100 \rangle$ and $\frac{1}{2}\langle 111 \rangle $ prismatic loops at 800~K are presented in subfigures \ref{fig:sia800K}(i) and \ref{fig:sia800K}(ii), respectively. Despite the simulations operating at high temperature on structures far from the training data, the simulations remain stable and well behaved. To smooth variations due to thermodynamic perturbations, the isosurfaces presented in (a) represent $\text{Tr}(\langle \sigma^k_{ij}(t)\rangle)$, where $\langle \sigma^k_{ij}(t)\rangle$ is the time averaged Virial stress over the simulation. (b) and (c) present contour maps of the time averaged stress ($\text{Tr}(\langle \sigma^k_{ij}(t)\rangle)$) and magnitudes of the MMs on a (100) or $(\bar{1}2\bar{1})$ plane cross section through the dislocation loops. Snapshots of the non-collinear MMs in and near the dislocation loops during dynamics at 800~K are shown in (d). The MMs near the defects become highly disordered relative to the bulk-like atoms. Importantly, the time averaged MMs shown in (e) for the tensile region of the dislocation are already acting paramagnetic despite the sample being below the Curie temperature. 

\section*{Discussion}

A new machine-learned spin-lattice potential (MSLP) for iron that can simultaneously simulate the mechanical and magnetic responses at finite temperature for both near-perfect and highly distorted configurations is developed. It is achieved through combining the knowledge of a conventional SLP and a neural network implemented using both local atomic and magnetic descriptors. Each MM is a three dimensional vector, where both the direction and magnitude depend on the local atomic environment and can be perturbed by thermal excitation.

Our MSLP shows near DFT accuracy on perfect crystals and point defect configurations. It produces quantitatively accurate predictions of various magnetic states in both BCC and FCC phases. The complex magnetic configurations in the vicinity of the core of vacancy and self-interstitial atom configurations, including the MM reversal and quenching at the core, were correctly reproduced. The order of stability of SIA configurations is compatible with DFT, where the $\langle 110\rangle $ dumbbell is most stable. 

Spin-lattice dynamics is performed to calculate the Curie temperature, which is in good agreement with experiment \cite{Lavrentiev_JPhysCM_2012}. We apply our potential to study the magnetism of mesoscopic scale dislocation loops at finite temperature. Non-collinear MMs about prismatic dislocation loops were investigated for the first time. We show moment magnitudes are suppressed in regions of compressive stress and are enhanced in regions of tensile stress. This transcends the capability of DFT and MD methods, as well as currently available MSLPs for iron. These simulations show good numerical stability at high temperature. Whilst the current MSLP is tailored to iron, the framework is flexible and can be applied to a large class of magnetic materials and alloys.

\section*{Methods}

\subsection*{Hamiltonian of machine-learned spin-lattice potential}\label{ssec:mlsld}

In many other developments of MP \cite{Behler_jcp_2011,Cooper_npj_2020,Nikolov_npj_2021,Novikov_natcompmat_2022}, the potential energy is defined as the output of a machine-learned machinery without any presumptions. The difference in the energy landscape can be up to the order of several eV, whilst requiring the accuracy and precision to be within at least $10^{-3}$ eV. Smoothness of the energy landscape is also a requirement of MP because atomic forces are calculated as the derivative of the potential energy. It necessitates a broad coverage of training data, especially near extrema. One can optimize their MP by supplying sufficient data to cover important parts of the phase space \cite{Novikov_natcompmat_2022} or generate massive amount of data in brute-force to cover the whole phase space \cite{Nikolov_npj_2021}.

On the other hand, if one can supply a \textit{mean function} before performing the learning process, the machine-learned machinery can then be used as a correction term. A properly chosen mean function can significantly reduce the amount of training data \cite{Rasmussen_Gaussian_Processes_2006}. We follow this logic and define our spin-lattice Hamiltonian as follows:
\begin{eqnarray}
\mathcal{H}(\mathcal{R},\mathcal{P},\mathcal{M}) =\sum_i \frac{\mathbf{p}^2_i}{2m}  + V(\mathcal{R},\mathcal{M}),
\end{eqnarray}
where
\begin{eqnarray}
V(\mathcal{R},\mathcal{M}) =V^{\text{NM}}(\mathcal{R}) + V^{\text{HL}}(\mathcal{R},\mathcal{M}) + V^{\text{NN}}(\mathcal{R},\mathcal{M}).
\end{eqnarray}
The Hamiltonian depends on the momenta $\mathcal{P}=\{\mathbf{p}_1,\mathbf{p}_2,...,\mathbf{p}_N\}$, atomic positions $\mathcal{R}=\{\mathbf{r}_1,\mathbf{r}_2,...,\mathbf{r}_N\}$ and magnetic moments $\mathcal{M}=\{\mathbf{M}_1,\mathbf{M}_2,...,\mathbf{M}_N\}$. The potential energy $V$ contains three terms. The non-magnetic term $V^{\text{NM}}$, which adopts the embedded atom method (EAM) functional form, takes care of the non-magnetic contributions. The Heisenberg-Landau (HL) term $V^{\text{HL}}$ takes care part of the magnetic contribution. The neural network (NN) term $V^{\text{NN}}$ takes care of contributions missed in $V^{\text{NM}}$ and $V^{\text{HL}}$. We may think its application as a correction term. The functional form of the first two terms follow conventional SLP \cite{ma_prb_2012,ma_prb_2017} that performs well near perfect crystal, but not in highly deformed configurations.

The NN term is defined as a sum of local contributions: 
\begin{equation}
    V^\text{NN} = V_0^\text{NN}\sum_i \mathcal{N}(\mathbf{x}^0_i;\{\mathbf{W},\mathbf{b}\}),
\end{equation}
where $V_0^{NN}$ is a fitting parameter to correct the energy unit. The neural network $\mathcal{N}$ is trained by adjusting the weight $\mathbf{W}$ and bias $\mathbf{b}$ parameters. The translation, rotation and permutation invariant atom centred symmetry (ACS) descriptors $\mathbf{x}^0_i$ of atom $i$ are extended to depend on both the local atomic and magnetic environments. This follows the usual assumption that the local environment is sufficient to determine the atomic energy \cite{Bartok_prb_2013}, where a cutoff distance $r_{\text{cut}}$ is adopted. 

Details of each term in the Hamiltonian are defined below. We performed fitting of all the parameters in $V$ using a massive amount of DFT data. The choice of Loss function, training data, training procedure, and final parameters are provided in the Supplementary Materials. In short, we developed a non-magnetic potential $V^{\text{NM}}$, followed by fitting parameters in $V^{\text{HL}}$ and $V^{\text{NN}}$. Finally, we optimized the whole potential $V$. One may consider $V^{\text{NM}}$ and $V^{\text{HL}}$ are used to construct a temporary mean function. On the contrary, each function in the NM and HL terms may be considered as a descriptor. As such, we can treat the $V$ as a special kind of machine-learned machinery that flexibly combines the known and unknown physics when trained to good quality data.

\subsection*{The Non-Magnetic term}\label{sssec:nm}

The Hamiltonian being adopted contains several terms, the non-magnetic term is chosen to have the same functional form of the embedded atom method (EAM) \cite{daw_msr_1993,Daw_prb_1984}:
\begin{equation}
    V^\text{NM}(\mathcal{R}) = \sum_i F(\rho_i) + \frac{1}{2}\sum_{i,j}V_{ij}(r_{ij}).
\end{equation}
$F(\rho_i)$ is a many-body term depending on the effective electron density $\rho_i$. $V_{ij}$ is a pairwise potential depending on $r_{ij}=|\mathbf{r}_i-\mathbf{r}_j|$, which is the distance between atom $i$ and $j$. The many-body term follows the functional form proposed by Mendelev\cite{Mendelev_philmag_2003} and Ackland\cite{Ackland_jpcm_2004}:
\begin{eqnarray}
F[\rho_i] = -\sqrt{\rho_i} + \phi \rho_i^2
\end{eqnarray}
where $\phi$ is a fitting parameter. The effective electron density $\rho_i$ is defined as a sum of the square of a pairwise function $t_{ij}$ which has the physical meaning corresponding to the hopping integral in tight binding model \cite{ma_prb_2017}:
\begin{equation}\label{eqn:eed}
    \rho_i = \displaystyle\sum_{j\in \{k | r_{ik} < r_{\text{cut}}\} } t_{ij}^2, 
\end{equation}
and
\begin{equation}
t_{ij}(r_{ij}) = \displaystyle\sum_n^{N_t}t_n(r^t_n-r_{ij})^3\Theta (r^t_n-r_{ij})
\end{equation}
where $t_n$ are fitting parameters, $r_n^t$ are knot points, and $\Theta$ is Heaviside function. The pairwise potential $V_{ij}$ is split into three parts:
\begin{eqnarray}
V_{ij} = 
\begin{cases}
V^{ZBL}(r_{ij}) & r_{ij}\le r_1 \\
V^{\text{it}}(r_{ij}) & r_1 < r_{ij} < r_2 \\
\displaystyle\sum_n^{N_V}V_n(r^V_n-r_{ij})^3\Theta (r^V_n-r_{ij}) & r_{ij}\ge r_2
\end{cases}
\end{eqnarray}
where $r_1 = 1.3${\AA} and $r_2 = 1.8${\AA}. The short-range part is ZBL potential \cite{zbl}. The middle-range part is a 5\tth order polynomial which ensures the function $V_{ij}$ being continuous up to second derivatives at $r_1$ and $r_2$. The longer-range part is a cubic spline, where $V_n$ are fitting parameters and $r_n^V$ are knot points. Numerical values of fitting parameters are mentioned in Supplementary Materials.

\subsection*{The Heisenberg-Landau term}\label{sssec:hl}

The Heisenberg-Landau term is a sum of a Heisenberg term $V^\text{H}$ and a Landau term $V^\text{L}$, such that
\begin{equation}
V^{\text{HL}}(\mathcal{R},\mathcal{M}) = V^{\text{H}}(\mathcal{R},\mathcal{M}) + V^{\text{L}}(\mathcal{R},\mathcal{M}).
\end{equation}

Conventional Heisenberg Hamiltonian assumes localised electron model with fixed magnitude of magnetic moments or spins \cite{ma_prb_2008,Perera_jap_2014,Tranchida_jcompphys_2018,Nikolov_npj_2021}. However, even for perfect crystalline configurations it has been observed that the adiabatic magnetic exchange-energy hypersurface parameterized by the bilinear Heisenberg Hamiltonian is insufficient \cite{Drautz2005,Okatov_prb_2011,Singer_prl_2011,Singer2011a}. An accurate representation necessitates longitudinal fluctuations to be considered \cite{Singer_prl_2011,Ruban_prb_2007,ma_prb_2012}, due to the itinerant nature of electrons. First, we write the Heisenberg Hamiltonian in a form that allows the change of magnitude \cite{Wang_prb_2010,chapman_prb_2020}:
\begin{eqnarray}\label{eqn:vh}
V^{\text{H}} = -\frac{1}{2}\displaystyle\sum_{ij}J_{ij}(r_{ij})\mathbf{M}_i\cdot \mathbf{M}_j.
\end{eqnarray}

The pairwise exchange coupling parameter $J_{ij}$ can be calculated through DFT according to the magnetic force theorem (MFT) \cite{Lichtenstein_jpfmp_1984}. We adopt a 5\tth order polynomial here that fits well to perfect BCC cases \cite{ma_prb_2017}:
\begin{equation}\label{eqn:jr}
    J(r_{ij}) = J_0 \left(1-\frac{r_{ij}}{r_{\text{cut}}}\right)^5\Theta (r_{\text{cut}}-r_{ij}).
\end{equation}

Second, the Heisenberg term can be improved by introducing higher order terms that describe longitudinal fluctuations \cite{chapman_prb_2020,Ruban_prb_2007,ma_prb_2012,rosengaard_prb_1997}. By using a Landau expansion, we introduce self-energy terms which create an energy well for a finite magnetic moment such that a spontaneous moment is formed and whose length can be variably controlled. We write the Landau term:
\begin{equation}
V^{\text{L}} = \sum_i\left(A(\rho_i)\mathbf{M}_i^2+B(\rho_i)\mathbf{M}_i^4+C\mathbf{M}_i^6\right).
\end{equation}
One can find more details regarding the philosophy of the Landau coefficients and how one may extract them directly from DFT calculations in Ref. \cite{ma_prb_2017, chapman_prb_2020}. Here, we simply treat them as fitting parameters. We assume an underlying quadratic polynomial functional form for both A and B coefficients, parameterised with respect to $\rho_i$ used in the EAM potential (Eqn.~\ref{eqn:eed}):
\begin{eqnarray}
    A(\rho_i) &=& a_0 + a_1 \rho_i + a_2 \rho_i^2, \\
    B(\rho_i) &=& b_0 + b_1 \rho_i + b_2 \rho_i^2.
\end{eqnarray}

The coefficient for the 6\tth order term is independent of the local environment and serves to prevent a divergence in the Landau energy well. It has been shown that such functional form is sufficient for strained on-lattice configurations, but is insufficient when lattice distortions are introduced \cite{chapman_prb_2020}. Further, DFT calculations have shown the magnitude of Landau coefficients in the core of defects can change by several orders of magnitude due to the suppression of the magnetic moments \cite{chapman_prb_2020}. As such, these terms provide an initial approximation. The neural network term serves as a necessary adjustment allowing the potential to move away from near-perfect crystal.

\subsection*{The neural network term}\label{sssec:mlc}

Conventional SLP is insufficient to reproduce the relative stability of BCC and FCC phases. MFT reveals the exchange coupling parameter $J_{ij}$ has completely different functional form for each crystal structures \cite{ma_prb_2017,chapman_prb_2020}. Previous work \cite{ma_prb_2017} defined two different set of $J_{ij}$ and Landau coefficients for each phase, such that the phase must be labelled \textit{a priori}, allowing free energy differences between the BCC and FCC phase to be extracted. It means such approach cannot be applied to arbitrary systems. A possible alternative is to calculate the $J_{ij}$ and Landau coefficients by DFT on the spot, but it is not feasible for large-scale atomic scale simulation. Besides, calculation of atomic Landau coefficient requires knowing the atomic energy, which is not a well defined quantity in most DFT implementations. 

To overcome the limitation imposed by the functional form, aiming at simulating arbitrary crystal structures, we apply machine learning techniques to develop a new potential. We choose artificial neural network (ANN) which abstractions between layers ensure magnetic interactions go beyond the bilinear form of the Heisenberg potential and local fluctuation of the Landau potential. 

Behler and Parinello\cite{BP_prl_2007} and others \cite{Behler_jcp_2011, Behler_jpcm_2014} successfully applied the ANN for atomic simulation based on feed-forward multilayer perceptrons. It composes of multiple layers of Threshold Logic Units (TLUs). They are fully connected between adjacent layers.  It is a feed forward ANN in which data provided to the input nodes are transmitted through one or more hidden layers until producing an output signal at the final layer. Unlike other architectures such as recurrent NN, cyclical connections between layers are not used. When two or more hidden layers are used, it is often referred to as a deep-ANN (DNN) \cite{wang_cpc_2018}. We simply call it NN in this work. 

In some works of MP for MD \cite{dragoni_prm_2018,Goryaeva_cms_2019,goryava_prm_2021,wang_cms_2022} a single machine-learned machinery, such as Gaussian process or NN, is used to predict the total energy, or more precisely the energy of an atom depending on the local atomic environment. Instead, we use NN to predict the contribution that cannot be captured by the non-magnetic term and Heisenberg-Landau term. In addition to the $V^\text{NM}$ and $V^\text{HL}$ term, the potential energy contains a NN term:
\begin{equation}
    V^\text{NN}=V_0^\text{NN}\sum_i \mathcal{N}(\mathbf{x}^0_i;\{\mathbf{W},\mathbf{b}\}),
\end{equation}
where $V_0^\text{NN}$ is a fitting parameter to match the scale and unit of the NN contribution to the MSLP Hamiltonian. $\mathbf{x}^0_i\in \mathbb{R}^{N_0}$ is a vector of descriptors with $N_0$ elements being supplied to the input layer of NN. Descriptors are functions depending on the atomic positions and magnetic moments within a cutoff distance $r_\text{cut}$ from atom $i$, representing the local atomic environment. The NN with $n$ layers is a mapping:
\begin{equation}
    \mathcal{N}(\mathbf{x}^0;\{\mathbf{W},\mathbf{b}\}) = \mathcal{P}^n \circ \mathcal{P}^{n-1} \circ \mathcal{P}^{n-2} \circ ... \circ \mathcal{P}^0(\mathbf{x}^0),
\end{equation}
where the operator $\circ$ represents the composition of functions. $\mathcal{P}^k$ is the mathematical description of a perceptron at layer $k$. It acts as a mapping from layer $k-1$ to the adjacent layer indexed $k$ ($0\leq k\leq n$) which includes the composition of a linear transformation, followed by a non-linear transformation using a component-wise activation function $\mathbf{f}_a^k$:
\begin{equation}
    \mathbf{x}^k = \mathcal{P}^k(\mathbf{x}^{k-1}) = \mathbf{f}_a^k(\mathbf{z}^k),
\end{equation}
where 
\begin{equation}
    \mathbf{z}^k = \mathbf{W}^k\mathbf{x}^{k-1} + \mathbf{b}^k. 
\end{equation}
$\mathbf{x}^k\in \mathbb{R}^{N_k}$ is a vector representation of the input signals from each of the $N_k$ nodes (neurons) in the k\tth layer of the NN. The weight matrix $\mathbf{W^k}\in \mathbb{R}^{N_k\times N_{k-1}}$ controls the strength of the signal transferred from each node in the $k-1$\tth layer to each node in the k\tth layer. $\mathbf{b}^k\in \mathbb{R}^{N_k}$ is a bias vector. The vector $\mathbf{z}^k$ is an intermediate quantity referred to as the weighted input. 

The activation function acts to abstractify the signals from the $k$\tth layer by adding non-linearity (since a linear combination of linear operations can itself be transformed into a single linear operation). It performs a component-wise operation on the weighted input $\mathbf{z}^k$ produced from the linear transformation of $\mathbf{x}^{k-1}$, such that
\begin{equation}
    \mathbf{f}_a^k(\mathbf{z}^k) = \left(f_a(z^k_0),f_a(z^k_1),...,f_a(z^k_{N_{k-1}})\right).
\end{equation}
We have chosen to use an unconventional unbounded activation function defined as:
\begin{eqnarray}
f_a(z) = \frac{z}{1+|z|}+az 
\end{eqnarray}
for all TLUs, except the output layer. The functional form of the activation function was chosen since the profile of $z/(1+|z|)$ is qualitatively similar to $\tanh (z)$ for small $z$ (approximately linear), but it is computationally cheaper than the hyperbolic tangent. Linear twisting is included to help prevent saturation for large values of $z$, which would result in a vanishing gradient of the Loss function, as originally proposed for the tanh function\cite{twisting}. The mapping of the final layer performs a linear transformation only, that is   $x^n=f^n_a(z^n)=z^n$ and produces a scalar output. Therefore, for $n$\tth layer the weight and bias terms have dimension $\mathbb{R}^{1\times N_{n-1}}$ and $\mathbb{R}^1$, respectively. 

The power of ANNs are due to their universality. For instance, a two-layer feed forward ANN with non-linear activation functions have been shown to be an universal function approximator. As such, for a given continuous function there exists a neural network which can approximate it on a compact set of $\mathbb{R}^N$ arbitrarily well \cite{Goodfellow-et-al-2016}. Furthermore, the universal approximator theorem has been shown to hold true for unbounded non-linear activation functions \cite{Sonoda_2017}. If linear activation functions were to be chosen, a DNN with any number of hidden layers may be represented as a single linear transformation and therefore cannot be a universal approximator.

\subsection*{Local atomic and magnetic descriptors}\label{sssec:descriptor}

In our MSLP, descriptors are functions representing the 6$N$ coordinate and spin space, where $N$ is the number of atoms. The purpose of the NN term is to map descriptors to part of the atomic energy. Since atomic energy is a scalar, descriptors should be translational, rotational and permutational invariant. We defined four sets of descriptors. The first set depends only on atomic positions. The other three depend on both atomic positions and magnetic moments.

Our atomic descriptors are based on the $G^{(2)}$ radial basis functions within the ACS class of effective coordinates \cite{Behler_jcp_2011,BP_prl_2007}. We drop the (2) superscript for brevity and refer to the descriptor as G2 in-text. It has been successfully used for a variety of materials including water \cite{Kondati_pccp_2015,Morawietz_jpc_2013}, aluminium and its alloys \cite{Kobayashi_prm_2017}, germanium telluride \cite{Sosso_prb_2012} as well as carbon allotropes \cite{Khaliullin_prb_2010}. It is written as:
\begin{equation}
    G_{i,h} \triangleq G^{(2)}_{i,h} = \sum_j G_{ij,h}, 
\end{equation}
where
\begin{equation}
   G_{ij,h} = w_i w_j f_c (r_{ij}) \exp \left( -\eta (r_{ij} - R_s)^2 \right),
\end{equation}
and $h$ is a compound index representing a unique triplet of hyperparameters $\{R_c,R_s,\eta\}$. The G2 descriptors are $R_s$ centred Gausssians spread according to $\eta$. $w_i$ and $w_j$ are weights which characterise different atomic species and are not uniquely defined. Often one maps a unique integer to each element type. Here we define them as the atomic mass (for iron $w=55.847$). Use of a species weight is advantageous as it enables descriptors to be defined which do not scale with the number of species. That is, the length of the input descriptor vector does not change with the number of chemical species.

The smoothness criterion is satisfied by employing an envelope function which, as well as its first derivative, decays smoothly to zero at the cutoff radius:
\begin{equation}
    f_c(r_{ij}) = 
    \begin{cases}
    \frac{1}{2}\bigg(\cos \bigg(\frac{\pi r_{ij}}{R_c}\bigg)+1\bigg) & r_{ij}\leq R_c \\
    0 & r_{ij}>R_c
    \end{cases}
\end{equation}

In this work we fix $R_c$ making it equals to the cutoff distance of the pair potential $r_{\text{cut}}$. It reduces the number of hyperparameters to 2. We used 9 equally spaced G2 descriptors with $R_s=2.0+x(R_c-2.0)/8$ where $x=0,1,2,...,9$. We set $\eta=5$ \AA$^{-2}$ to provide a small overlap between the Gaussian basis functions. Often a large number of G2 descriptors (5-200) are used varying $\eta$ from $10^{-2}$ to $1$ \AA$^{-2}$ with $R_s=0$ \AA \cite{Goryaeva_cms_2019}. We opted to fix the Gaussian width and varying their centering, in order to reduce the correlation between the data encoded by each descriptor. Reducing the correlation can also be achieved using more advanced orthogonal descriptors such as SOAP\cite{Bartok_prb_2013} at the expense of greater computation time per descriptor. Since we consider $6N$ degrees of freedoms, we chose G2 descriptor as it is computationally less demanding.

The design of the magnetic descriptors is based on the G2 function. Inspired by the Heisenberg and Landau functional forms, we write three further sets of descriptors. The first set of magnetic descriptors is written as:  
\begin{equation}
G^{H}_{i,h} = \displaystyle\sum_j G^{H}_{ij,h},
\end{equation}
where the 2-body contributions are defined as:
\begin{equation}
G^{H}_{ij,h} \triangleq G_{ij,h}\mathbf{M}_i \cdot \mathbf{M}_j .
\end{equation}
The scalar product of the magnetic moments ensures the invariant properties are maintained. Smoothness is guaranteed by the G2 prefactor which contains the envelope function $f_c$. By reusing the G2 in the magnetic descriptors we reduce the computational cost of the descriptor calculations which must be performed for every atom, at every timestep if dynamics is to be performed. Each G2 in the Heisenberg-like descriptors may be considered to be surrogate exchange parameters with different dependencies on the local environment as set by the chosen hyperparameters.

Similarly, we defined descriptors inspired by the Landau term up to the 4\tth order. A 6\tth order term is provided in the classical Landau expression to prevent divergences. As with the Heisenberg-like term, the Landau-like descriptors are built from a sum of two body contributions:
\begin{eqnarray}
    G^{A}_{ij,h}   &\triangleq& G_{ij,h}\mathbf{M}_i^2,\\
    G^{B}_{ij,h}   &\triangleq& G_{ij,h}\mathbf{M}_i^4.
\end{eqnarray}
Each hidden layer of NN provides successively higher order representations of the exchange interactions beyond the original bilinear, quadratic and quartic input interactions. 

We use nine G2 radial basis functions. The input of the NN has a dimension of $\mathbb{R}^{N_0}$, where $N_0=4\times 9=36$. That is, nine structural descriptors $\{G^{(2)}\}$, nine Heisenberg-like descriptors $\{G^{H}\}$, nine Landau-A-like descriptors $\{G^{A}\}$ and nine Landau-B-like descriptors $\{G^{B}\}$. Every descriptor has analytical derivatives with respect to both changes in position and magnetic moment (see Supplementary Materials). Whilst we opted to use two-body G2 as the basis of our magnetic descriptors, the principle is extendable to N-body descriptor representations.

\subsection*{Fitting Procedure}

Once the database has been constructed, the model parameters can be trained by minimising the Loss function (see Supplementary Materials). Each component in the model Hamiltonian is motivated by different physical properties. To reflect this our fitting workflow consisted of four distinct stages to allow each term to learn their respective physical behaviours.

\begin{enumerate}
    \item First, we introduce the underlying behaviour of metallic bonds for BCC, FCC and HCP iron in the absence on magnetic interactions by fitting the parameters of the non-magnetic potential $V^{\text{NM}}(\mathbf{p}^{\text{NM}})$ to the configurations in the non-magnetic database. The non-magnetic parameters are the subset $\mathbf{p}^{\text{NM}}=(\{V^t\},\{r^t\},\{t^N\},\{r^t\},\phi)$. We maintain the parameters of the ZBL potential. The coefficients of the interpolation potential $V^{it}$ are not fit but are analytically derived to maintain continuity.
    \item Next, the characteristic behaviour of \textit{band splitting} (i.e. the spontaneous formation of a magnetic moment) and their itinerant magnetic interactions are added by fitting the Heisenberg-Landau parameters $\mathbf{p}^{\text{HL}}$ to bulk-like BCC configurations in the magnetic database. During this process $\mathbf{p}^{\text{NM}}$ are held constant such that the total energy considered by the loss function is $\mathcal{U}\to V^{\text{NM}}+V^{\text{MC}}(\mathbf{p}^{\text{HL}}$). The Heisenberg-Landau parameters are $\mathbf{p}^{\text{HL}}=(J_0,\{a\},\{b\},c)$.
    \item Magnetic interactions beyond the parametric constraints of the Heisenberg-Landau formalism are produced by training the NN weights and biases $\mathbf{p}^{\text{NN}}=(\{\mathbf{W}\},\{\mathbf{b}\},V_0)$ to all desired observables in the magnetic database. This also introduces the magnetic behaviour of the FCC phase into the Hamiltonian. During this stage the total energy is given by the full model $\mathcal{U}\to V^{\text{NM}}+V^{\text{HL}}+V^{\text{NN}}(\mathbf{p}^{\text{NN}}$), where the parameters $\mathbf{p}^{\text{NM}}$ and $\mathbf{p}^{\text{HL}}$ are fixed.
    \item Finally, we make minor adjustments to the parameter space by enabling all variables $\mathbf{p}=(\mathbf{p}^{\text{NM}},\mathbf{p}^{\text{HL}},\mathbf{p}^{\text{NN}})$ to be simultaneously adjusted with total energy: $\mathcal{U}\to V^{\text{NM}}(\mathbf{p}^{\text{NM}})+V^{\text{HL}}(\mathbf{p}^{\text{HL}})+V^{\text{NN}}(\mathbf{p}^{\text{NN}}$). In this stage the maximum step size of the minimization algorithm is reduced.
\end{enumerate}

Once the Loss function has been minimised with respect to the training database its generalisation and stability may be validated through dynamic simulations. This is usually performed using MD. However, our model Hamiltonian has coupled spin and lattice degrees of freedoms and is designed to incorporate itinerant behaviour. Consequently, in order to evolve with time and temperature we may use SLD to treat atomic and magnetic interactions on equal footing which already incorporated both transverse and longitudinal magnetic fluctuations.

\bibliography{Fe_MLSLD}

\section*{Acknowledgements}

We acknowledge Sergei L. Dudarev for stimulating discussion. 
This work has received funding from the Euratom research and training programme
2014-2018 under grant agreement No. 755039 (M4F project).
This work has been carried out within the framework of the EUROfusion Consortium, funded by the European Union via the Euratom Research and Training Programme (Grant Agreement No 101052200 — EUROfusion) and from the EPSRC [grant number EP/T012250/1]. Views and opinions expressed are however those of the author(s) only and do not necessarily reflect those of the European Union or the European Commission. Neither the European Union nor the European Commission can be held responsible for them. This work has been part-funded by the EPSRC Energy Programme [grant number EP/W006839/1]. To obtain further information on the data and models underlying this paper please contact PublicationsManager@ukaea.uk.
We acknowledge EUROfusion for the provision of access to Marconi-Fusion HPC facility. The authors acknowledge the use of the Cambridge Service for Data Driven Discovery (CSD3) and associated support services provided by the University of Cambridge Research Computing Services (www.csd3.cam.ac.uk) in the completion of this work.

\section*{Author contributions statement}

PWM conceived the original MSLD method with contributions from JBJC. JBJC and PWM developed the MSLD training software. JBJC performed the potential fitting. JBJC implemented the MSLP into the SPILADY code originally written by PWM. Both JBJC and PWM analysed the results, wrote and reviewed the manuscript.

\section*{Additional information}

To include, in this order: \textbf{Accession codes} (where applicable); \textbf{Competing interests} (mandatory statement). 

The corresponding author is responsible for submitting a \href{http://www.nature.com/srep/policies/index.html#competing}{competing interests statement} on behalf of all authors of the paper. This statement must be included in the submitted article file.

\end{document}